\documentclass[lettersize,journal]{IEEEtran}
\usepackage{amsmath,amsfonts}
\usepackage{algorithmic}
\usepackage{array}
\usepackage{ntheorem}
\usepackage{textcomp}
\usepackage{stfloats}
\usepackage{url}
\usepackage{verbatim}
\theoremstyle{plain}
\theorembodyfont{\unshape}

\usepackage{graphicx}
\usepackage{mathrsfs}
\usepackage{subfigure}
\usepackage{multirow}
\usepackage{algorithm}
\usepackage{enumerate}
\usepackage{color}
\usepackage[numbers,sort&compress]{natbib}

\def\BibTeX{{\rm B\kern-.05em{\sc i\kern-.025em b}\kern-.08em
    T\kern-.1667em\lower.7ex\hbox{E}\kern-.125emX}}
\usepackage{balance}
\begin{document}

\title{Deep Reinforcement Learning-Assisted Federated Learning for Robust Short-term Utility Demand Forecasting in Electricity Wholesale Markets}

\author{Chenghao Huang, Weilong Chen,
Shengrong Bu~\IEEEmembership{Member,~IEEE}, Yanru Zhang~\IEEEmembership{Member,~IEEE}
}





\maketitle



\begin{abstract}
Short-term load forecasting (STLF) plays a significant role in the operation of electricity trading markets. Considering the growing concern of data privacy, federated learning (FL) is increasingly adopted to train STLF models for utility companies (UCs) in recent research. Inspiringly, in wholesale markets, as it is not realistic for power plants (PPs) to access UCs' data directly, FL is definitely a feasible solution of obtaining an accurate STLF model for PPs. However, due to FL's distributed nature and intense competition among UCs, defects increasingly occur and lead to poor performance of the STLF model, indicating that simply adopting FL is not enough. In this paper, we propose a DRL-assisted FL approach, DEfect-AwaRe federated soft actor-critic (DearFSAC), to robustly train an accurate STLF model for PPs to forecast precise short-term utility electricity demand. Firstly. we design a STLF model based on long short-term memory (LSTM) using just historical load data and time data. Furthermore, considering the uncertainty of defects occurrence, a deep reinforcement learning (DRL) algorithm is adopted to assist FL by alleviating model degradation caused by defects. In addition, for faster convergence of FL training, an auto-encoder is designed for both dimension reduction and quality evaluation of uploaded models. In the simulations, we validate our approach on real data of Helsinki's UCs in 2019. The results show that DearFSAC outperforms all the other approaches no matter if defects occur or not.
\end{abstract}



\let\thefootnote\relax\footnotetext{C. Huang, W. Chen, and Y. Zhang are with the College of Computer Science and Engineering, University of Electronic Science and Technology of China, Chengdu, China, 611731 (e-mail: zydhjh4593@gmail.com, chenweilong1995@std.uestc.edu.cn, yanruzhang@uestc.edu.cn). Y. Zhang is the corresponding author.}
\let\thefootnote\relax\footnotetext{S. Bu is with the Department of Engineering, Brock University, St. Catharines, Canada, L2S 3A1 (e-mail: sbu@brocku.ca).}

\section{Introduction}\label{intro}
In recent years, many countries and regions have gradually opened up their electricity trading markets, in which utility companies (UC) purchase electricity from power plants (PP) in a wholesale market, and then sell it to consumers in a retail market. As the number of UCs increases, Texas gradually occupies the largest share of the electricity trading market in the US. By November $2021$, more than $110$ UCs have been operating in the Electric Reliability Council of Texas \cite{plan}. As power supply and demand have a significant influence on energy transactions, market shares, and profits in competitive electricity markets, a precise load forecast, especially short-term load forecasting (STLF), is essential for electricity price estimation \cite{ceperic2013strategy}. 

In the smart grid, with the widespread deployment of advanced metering infrastructure (AMI) in various buildings, such as residential buildings, commercial buildings, industrial buildings, etc., approaches of STLF on electricity have been actively studied. The work of \cite{kong2017short} adopts long short-term memory (LSTM) networks to extract temporal features in electricity consumption data of residential buildings, which has become a popular way in STLF. Based on \cite{kong2017short}, research on STLF using deep learning (DL) springs up. As UCs can collect abundant consumer profiles, such as historical load data, household characteristics, and behavior patterns through AMIs, STLF of residential buildings and communities is well developed \cite{oprea2019machine,9144528,afrasiabi2020deep,hong2020deep,lin2021spatial}. In addition, there also exist some DL-based STLF approaches on industrial load data \cite{bracale2017short, ahmad2016accurate, wang2020short}, commercial load data \cite{chitalia2020robust}, hospital load data \cite{9621428}, and so on.

To maintain the stability of electricity trading markets, STLF on UCs' demand is also necessary for PPs. However, as there is a great competition of pricing in wholesale markets in Texas \cite{texus}, it is not realistic for PPs to gain UCs' data. Besides, PPs have no AMIs or any other devices to access consumers' profiles. Therefore, an approach of obtaining an accurate STLF model for PPs without intruding UCs' data privacy is strongly needed. 


Considering the growing concern of protecting data privacy, federated learning (FL) \cite{li2020federated}, which aims at providing general solutions while ensuring data privacy and security, is adopted in most research on STLF \cite{taik2020electrical,savi2021short}, indicating the feasibility of adopting FL between one PP and UCs to help the PP obtain an accurate STLF model.

Nevertheless, as mentioned above, the competition in wholesale markets is intense, implying that malicious UCs inevitably exist. Due to FL's distributed nature, bad conditions increasingly occur in FL-based STLF model training on PPs. In this paper, we call all bad conditions \textbf{defects}. As stated in research \cite{qureshi2022poisoning}, malicious hackers or UCs can conduct various attacks, such as data poisoning attacks, to training data or trained models. Besides, uneven quality of communication can also introduce errors during uploading local models or downloading the global model. Thus, a robust design is eagerly required to be proposed, rather than simply adopting FL. 


To tackle the challenge, deep reinforcement learning (DRL) \cite{sutton2018reinforcement} is adopted. In DRL, an agent is trained to interact with the environment, which has the strong capability of solving real-time decision tasks with significant uncertainty. As DRL becomes increasingly prevalent in solving problems which can be modeled as Markov decision process (MDP) \cite{puterman1990markov}, many studies of adopting DRL in smart grid emerge \cite{glavic2019deep,park2020short,feng2019reinforcement}. On the other hand, the work of \cite{wang2020optimizing} combines FL and DRL by selecting a certain number of clients from all clients through a DRL model to deal with image classification tasks. Motivated by the above works, we adopt a DRL algorithm to output the optimal weights for uploaded models so that the PP can aggregate model parameters to the global model which can be guaranteed an improvement in each round despite defects occurring.

As we know, there are just few works about adopting FL in probabilistic solar load forecasting \cite{zhang2020probabilistic, lin2021privacy}. Considering the mature development and wide deployment of deterministic load forecasting (DLF) in demand, we still construct STLF model based on DLF in this paper.

The main contributions are summarized as follows:
\begin{itemize}
    \item \textit{An approach of obtaining a STLF model for PPs without intruding data privacy}: As far as we know, there is no work of adopting FL to obtain a STLF model for PPs. In this paper, FL is adopted to aggregate UCs' STLF models to a global model with high accuracy for PPs. In return, UCs can download the global model for better local STLF. Above all, the data privacy of UCs in wholesale markets is protected through FL.
    \item \textit{A DRL-based design for robustness against defects}: Considering defects occurring in wholesale markets, uploaded models, which have different quality, may harm the performance of the global model during model aggregation on the PP server. To alleviate the model degradation caused by defects, a DRL algorithm, soft actor-critic (SAC), is adopted to assign optimal weights to uploaded models to guarantee efficient model aggregation, which makes the FL process significantly robust.
    \item \textit{Model dimension reduction and quality evaluation}: Since high-dimensional model parameters are uploaded to the server, as well as the assigned weights are continuous, only relying on DRL leads to massive time and computational resources for convergence. Thus, inspired by some techniques of dealing with the defects in FL \cite{tuor2021overcoming,yang2022robust}, an auto-encoder, called quality evaluation embedding network (QEEN), is constructed before the DRL model to reduce the dimension of uploaded models and evaluate their quality to accelerate the DRL training.
\end{itemize} 

To sum up, a DRL-assisted FL approach, named DEfect-AwaRe federated soft actor-critic (DearFSAC), is proposed to robustly integrate an STLF model for PPs using UCs' local models.

The remaining parts of this paper proceed as follows. In Section \ref{sec3}, we formulate the communication between one PP and UCs under an FL paradigm. DearFSAC is elaborated in Section \ref{sec4}. In Section \ref{sec5}, we evaluate our approach through simulations. Finally, we conclude our work in Section \ref{sec6}.

\section{System Model}\label{sec3}
\subsection{FL between one PP and UCs}
\begin{figure}[ht]
\centering
\includegraphics[width=.46\textwidth]{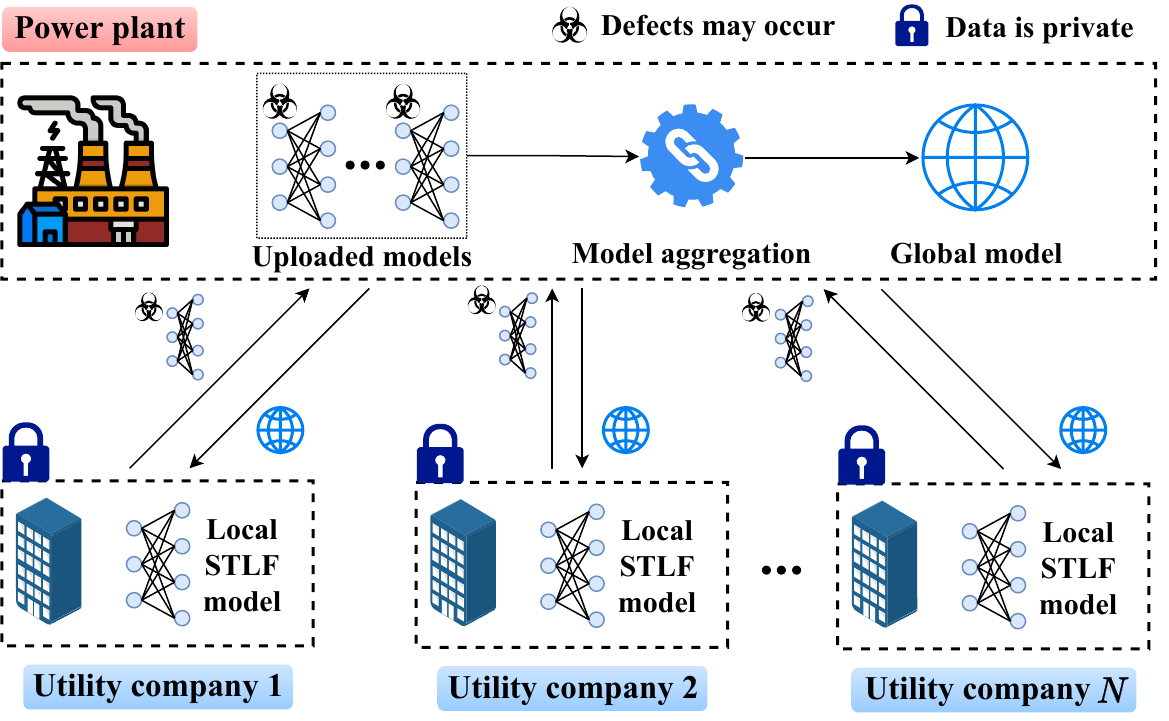}
\caption{The FL framework of model aggregation for one PP using UCs' models with defects occurring.}
\label{system model}
\end{figure}

Assume there are $N$ UCs and one PP. The UCs are set as clients who upload their models to the server, i.e. the PP. Firstly, clients upload their STLF models. Then, the server aggregates uploaded local models to the global model and allows clients to download the global model for further local training. The two steps form a loop, illustrated in Fig. \ref{system model}.

However, during the FL process, various defects may occur and decrease the accuracy of the PP's STLF model. Briefly, in this paper, we consider the following $2$ defects:
\begin{itemize}
    \item \textit{Data integrity attacks (DIAs)}: As the quality of input data affects the forecast accuracy directly, DIAs, where hackers access supposedly protected data and inject false information, are harmful to load forecasting and hard to be detected directly \cite{luo2018benchmarking}. In FL, DIAs weaken both local models and the global model.
    \item \textit{Communication noises}: Due to the frequent transmission of model parameters between UCs and the PP \cite{qureshi2022poisoning}, the uploaded parameters can be inaccurate, which randomly weaken the performance of model aggregation.
\end{itemize} 

Two issues should be addressed in this FL framework:
\begin{itemize}
    \item Since UCs' data privacy is concerned in wholesale markets, the PP's STLF model should be designed to well capture the hidden temporal features only from the historical load data and time data.
    \item Since various defects occur in FL process, a robust model aggregation approach is needed to obtain a global model with high STLF accuracy.
\end{itemize}


\begin{table}
\caption{Summary of Main Notations}
\centering
\begin{tabular}{cp{7cm}}
\hline
Notation  & Meaning \\
\hline
$\mathbf{a}$    &    The weight vector\\
$C_i$  &    The $i$th UC \\
$\mathcal{D}_i$    &    The electricity consumption dataset owned by $C_i$\\
$f_i(\cdot)$    &    The loss of a model evaluated on $\mathcal{D}_i$  \\
$F(\cdot)$    &    The objective of STLF adopting FL with defects occurring  \\
$N$    &    Total number of UCs in wholesale market  \\
$p_K$    &    The proportion of selecting uploaded models in each round\\
$p_M$    &    The proportion of defective models among all clients  \\
$T_i$ & The time period of $\mathcal{D}_i$ \\
$\mathbf{w}^i$    &   Local model parameters of $C_i$ \\
$\mathbf{w}^g$    &   Global Model parameters of the server\\
$\Tilde{\mathbf{w}}$    &   Defective local model\\
$\mathbf{X}_i$    &    Features of $\mathcal{D}_i$  \\
$\mathbf{y}_i$    &    True electricity consumption of $\mathcal{D}_i$  \\
$\hat{\mathbf{y}}_i$    &    Predicted electricity demand of $C_i$  \\


\hline
\end{tabular}
\label{notation}
\end{table}


\subsection{Formulation}\label{Problem formulation}
The main notations of this work are summarized in Table I. $N$ clients and corresponding raw datasets are defined in Definition 2.1:

\textit{Definition 2.1:} Let $\{C_i\}^N_{i=1}$ be the set of $N$ clients of an FL process, and $C_i$'s dataset is $\mathcal{D}_i=<\mathbf{X}_i,\mathbf{y}_i>$, where $\mathbf{X}_i=\{\mathbf{x}^1_i,...,\mathbf{x}^{T_i}_i\}$ is the historical electricity consumption and time data, and $\mathbf{y}_i=\{y^1_i,...,y^{T_i}_i\}$ is a vector containing the true electricity demand. For each client $C_i$, $\hat{\mathbf{y}}_i=\{\hat{y}^1_i,...,\hat{y}^{T_i}_i\}$ represents a vector containing the predicted demand of $C_i$. To conduct model aggregation, the STLF models of $\{C_i\}^N_{i=1}$ are with the same structure. Then, the parameters of $C_i$ and the server are represented by parameter vectors $\mathbf{w}^i \in \mathbb{R}^{d}$ and $\mathbf{w}^g \in \mathbb{R}^{d}$, respectively, where $d$ is the total number of one model's parameters. Thus, as $\mathbf{w}^g$ is downloaded by all clients, and $\mathbf{X}_i$ is inputted without any preprocess, $\hat{\mathbf{y}}_i$ can be computed as:
\begin{align}
    \hat{\mathbf{y}}_i = \text{STLF}(\mathbf{w}^g, \mathbf{X}_i).
\end{align}





Based on Definition 2.1, to minimize the averaged loss of testing $\mathbf{w}^g$ on $\mathcal{D}_i$, the objective of FL can be formulated into an empirical risk minimization (ERM) problem as follows:
\begin{align}
     \label{fl goal} \min_{\mathbf{w}^g \in \mathbb{R}^{d}} \Big[F(\mathbf{w}^g) = \frac{1}{N|\mathcal{D}_i|} \sum^{N}_{i=1}f_{i}(\mathbf{w}^g)\Big],
\end{align}
where $\mathbf{w}^g$ is downloaded by clients, and $f_i(\mathbf{w}^g)$ represents the loss of the model $\mathbf{w}^g$ trained on $\mathcal{D}_i$, formulated as:
\begin{align}
    f_i(\mathbf{w}^g)&=\mathbb{E}_{\xi_i \sim \mathcal{D}_i}\Big[\frac{1}{|\xi_i|} \sum^{|\xi_i|}_{t=1} (y^{\xi_{i,t}}_i-\hat{y}^{\xi_{i,t}}_i)^2 \Big], \quad y^{\xi_{i,t}}_i \in \xi_i.
\end{align}
where $\xi_i=\{y^{\xi_{i,1}}_i,...,y^{\xi_{i,|\xi_i|}}_i\}$ is a subset of $\mathcal{D}_i$.

For the server, the objective is to find the optimal global model parameters:
\begin{equation}
    \mathbf{w}^{g*} = \mathop{\arg \min}_{\mathbf{w}^g \in \mathbb{R}^{d}} {F(\mathbf{w}^g)}.
\end{equation}

For better performance, at each round, $K$ models among all uploaded models are randomly selected as a subset $\mathbb{N}_K=\{k_1,...,k_K\}$ to participate in model aggregation, where $K=\lfloor p_K N \rfloor$ and $p_K$ is a proportion within $[0,1]$ \cite{wang2020optimizing}. Assume we have a weight vector $\mathbf{a}=(a^1,...,a^K)$ outputted by the DRL model. The PP multiplies $\mathbf{a}$ and all selected model parameters $\mathbf{W} = (\mathbf{w}^{k_1},...,\mathbf{w}^{k_K})$ to get the global model:
\begin{equation}
    \mathbf{w}^g = \mathbf{a}\mathbf{W} = \sum^{K}_{i=1} a^i \mathbf{w}^{k_i}.
\end{equation}

Next, considering that defects occur in FL process or smart grid, we define defective models in Definition 2.2:

\textit{Definition 2.2:} Assume there is a subset $\mathbb{N}_M=\{m_1,...,m_{\lfloor MN \rfloor}\}$ contains $M$ models among all uploaded models affected by defects, where $M=\lfloor p_M N \rfloor$ and $p_M$ is a proportion within $[0,1]$. If $\mathbf{w}^i$ is defective, it can be denoted as $\Tilde{\mathbf{w}}^i$. In each FL communication round, we assume there is a subset of defective models among selected models, denoted as $\mathbb{K}_M=\mathbb{N}_M \cap \mathbb{N}_K$. 



\begin{figure*}[ht]
\centering
\includegraphics[width=.95\textwidth]{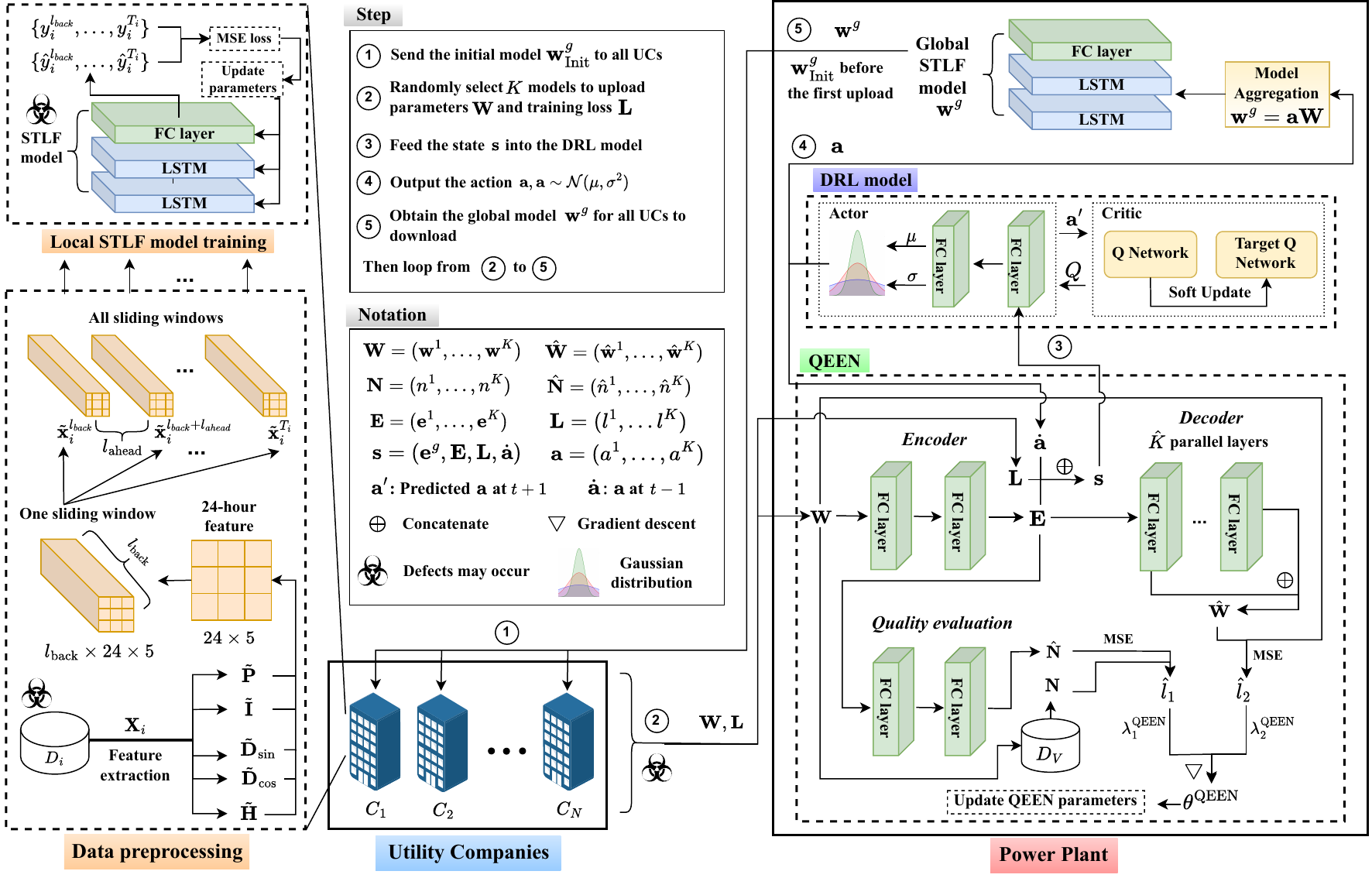}
\caption{The overall architecture of DearFSAC, in which the LSTM forecasting model is introduced in Subsection \ref{local training}, QEEN is introduced in Subsection \ref{QEEN-sec}, and the DRL model is introduced in Subsection \ref{SAC in FL}.}
\label{DearFSAC}
\end{figure*}

The framework mentioned above can be illustrated in Fig. \ref{system model}. As $f_i(\Tilde{\mathbf{w}}^i)$ is much higher than $f_i(\mathbf{w}^i)$, and it is uncertain which models are defective, fixed weights, such as averaged ones, will conduct poor model aggregation. Therefore, an optimal weights assignment approach is required. The objective of FL model aggregation with defects occurring can be formulated as:
\begin{align}
    \min & F(\mathbf{a}\mathbf{W}), \\
    F(\mathbf{a}\mathbf{W})&=\frac{1}{N}\sum^{N}_{i=1}f_i(\sum^{K}_{j=1} a^j \mathbf{w}^{k_j}),\\
s.t. & \sum^{K}_{j=1} a^j = 1, \notag\\
     & j \in \{1,...,K\}, \notag
\end{align}
where if ${k_j}$th model is in $\mathbb{K}_M$, it is a defective model $\tilde{\mathbf{w}}^{k_j}$. Then, an optimal weight vector $\mathbf{a}^*$ is needed to be found:
\begin{align}
    \mathbf{a}^* &= \mathop{\arg\min}_{\mathbf{a}} \Big[F(\mathbf{a}\mathbf{W})\Big].
\end{align}


\section{Proposed DEfect-AwaRe Federated Soft Actor-critic Approach}\label{sec4}


\subsection{Overall Architecture}\label{overview}
A novel FL-based framework, DearFSAC, is proposed to accurately forecast the total electricity demand of UCs for the PP to generate appropriate electricity while defects occur. The overall architecture of DearFSAC is shown in Fig. \ref{DearFSAC}. In general, this approach mainly consists of $3$ modules, which are: 1) STLF model based on LSTM; 2) The quality evaluation embedding network (QEEN); 3) The DRL model based on soft actor-critic (SAC). As the PP just has historical data and time data, the STLF model should be capable of capturing hidden temporal features. Furthermore, during the FL process, considering different quality of uploaded models and various defects, the DRL model based on SAC is adopted to assign optimal weights to uploaded models to conduct efficient aggregation. Besides, just inputting model parameters into the DRL model will lead to curse of dimensionality and quite slow convergence. Therefore, QEEN is designed to reduce uploaded model parameters' dimension and evaluate these models' quality to provide more effective information for faster convergence of the DRL model.

\subsection{STLF model based on LSTM}\label{local training}

Based on the work from \cite{kong2017short}, we adopt the LSTM \cite{hochreiter1997long} network as the structure of the STLF model. In order to make the LSTM model work, the inputs need to be time series. Therefore, we create sliding windows with $l_{\text{back}}$ and $l_{\text{ahead}}$ from the data, where $l_{\text{back}}$ and $l_{\text{ahead}}$ are the length in days of one sliding window and the interval of sliding windows' starting points, respectively. In our work, each daily consumption profile consists of $24$ hourly intervals, and thus the width of all feature vectors should be $24$. After data cleaning and feature selection, input features containing temporal information are generated, which include:
\begin{itemize}
    \item The sequence of electricity consumption for the past $\tilde{T}$ time intervals $\mathbf{P}\in\mathbb{R}^{\tilde{T}}$.
    \item The incremental sequence of the hourly time indices for the past $\tilde{T}$ time steps $\mathbf{I}\in\mathbb{R}^{\tilde{T}}$.
    \item The corresponding day of a week is within $\{0,...,6\}$, which is simply mapped to $\tilde{d}_{\text{sin}}\in[-1,1]$ and $\tilde{d}_{\text{cos}}\in[-1,1]$ using trigonometric functions.
    \item The binary holiday marks of corresponding dates $\tilde{h}$ are within $\{0,1\}$.
\end{itemize}

To ensure the feature dimensions are consistent, we extend $\tilde{d}_{\text{sin}}$, $\tilde{d}_{\text{cos}}$, and $\tilde{h}$ to 24-dimensional vectors $\tilde{\mathbf{D}}_{\text{sin}}=(\tilde{d}_{\text{sin}},...,\tilde{d}_{\text{sin}})$, $\tilde{\mathbf{D}}_{\text{cos}}=(\tilde{d}_{\text{cos}},...,\tilde{d}_{\text{cos}})$, and $\tilde{\mathbf{H}}=(\tilde{h},...,\tilde{h})$, respectively. After we normalize $\mathbf{P}$ and $\mathbf{I}$ to $\tilde{\mathbf{P}}$ and $\tilde{\mathbf{I}}$, respectively, where elements in $\tilde{\mathbf{P}}$ and $\tilde{\mathbf{I}}$ are both within $[0,1]$, the input of the LSTM model can be constructed into a matrix of the five transposed vectors:
\begin{equation}
    \mathbf{x} = \{\tilde{\mathbf{P}} ^ \top,\tilde{\mathbf{I}} ^ \top, \tilde{\mathbf{D}}_{\text{sin}} ^\top, \tilde{\mathbf{D}}_{\text{cos}} ^\top, \tilde{\mathbf{H}}^\top\}.
\end{equation}

As $\tilde{\mathbf{P}} ^ \top$, $\tilde{\mathbf{I}} ^ \top$, $\tilde{\mathbf{D}}_{\text{sin}} ^\top$, $\tilde{\mathbf{D}}_{\text{cos}} ^\top$, and $\tilde{\mathbf{H}}^\top$ are all one-dimensional vectors, the dimension of a sliding window is $l_{\text{back}}\times 24 \times 5$. After the above preprocessing, the training input of $C_i$'s forecasting model is a set of sliding windows, which can be represented by $\tilde{\mathbf{X}}_i=\{\tilde{\mathbf{x}}^{l_{\text{back}}}_i,\tilde{\mathbf{x}}^{l_{\text{back}}+l_{\text{ahead}}}_i...,\tilde{\mathbf{x}}^{T_i}_i\}$, in which $\tilde{\mathbf{x}}^{l_{\text{back}}}_i$ is the first sliding window computed from $l_{\text{back}}$th time step. To update LSTM model parameters, mean square error (MSE) is adopted as the loss function. Then at FL round $t$, the parameters $\mathbf{w}^i_t$ can be updated as follows:
\begin{align}
    \mathbf{w}^i_{t+1} &= \mathbf{w}^i_t - \alpha_{\text{LSTM}}\bigtriangledown l_t^i,\\
    l_t^i &= \text{MSELoss}(\hat{\mathbf{y}}_i, \mathbf{y}_i),
\end{align}
where $\mathbf{w}^i_{t+1}$ are the updated model parameters, $l_t^i$ is the training loss, and $\alpha_{\text{LSTM}}$ is the learning rate of local training.

When using this model, at time $\tilde{t}$, the input is a sliding window counting back $l_{\text{back}}$ time steps from $\tilde{t}$, and output of is the hourly electricity demand at time $\tilde{t}$, where $\tilde{t}$ is the time index in the dataset $\mathcal{D}_i$.



\subsection{Dimension Reduction and Quality Evaluation via QEEN}\label{QEEN-sec}
For quality evaluation and dimension reduction, QEEN, an auto-encoder is introduced before the DRL model. For training efficiency of QEEN, we upload all local model parameters $\mathbf{W}=(\mathbf{w}^1,...,\mathbf{w}^N)$ to the server. In other words, we set $p_K=1$. Then we design loss $\hat{l}_1$ for the embedding of $\mathbf{w}^i$ and loss $\hat{l}_2$ for quality prediction.



Firstly, we define the embedding vectors in Definition 3.1:

\textit{Definition 3.1:} We feed each $\mathbf{w}^i \in \mathbf{W}$ into the encoder $f_{\text{Enc}}(\cdot)$ composed of two FC layers to get the embedding vector $\mathbf{e}^i$ of the $i$th model. After obtaining all embedding vectors, we put the embedding vector concatenation $\mathbf{E}=(\mathbf{e}^1,...,\mathbf{e}^N)$ into the decoder $f_{\text{Dec}}(\cdot)$ to produce a decoded representation $\hat{\mathbf{W}}$ which approximates $\mathbf{W}$.

For faster training, we design the decoder into parallel FC layers $f_{\text{Dec}}=(f^1_{\text{Dec}},...,f^{\hat{K}}_{\text{Dec}})$, where $f^k_{\text{Dec}}$ is the $k$th parallel FC layer corresponding to the $k$th layer of the original model structure \cite{zinkevich2010parallelized}, and $\hat{K}$ is the number of layers of the original model. Next, for the $i$th model, the embedding vector $\mathbf{e}^i$ is fed into the $k$th parallel layer to get decoded layer parameters $\mathbf{w}^{\{i,k\}}$ of the original model. By concatenating each $\mathbf{w}^{\{i,k\}}$ layer by layer, the entire decoded model parameters $\hat{\mathbf{w}}^i$ are obtained.


As multiple defects have different impacts on local models, we define defect marks and quality evaluation marks in Definition 3.2:

\textit{Definition 3.2:} To train the QEEN for quality evaluation, we set the defect marks $\mathbf{N}=(n^1,...,n^N)$ as the ground truth, where $n^i$ represents the severity of defects in $\mathbf{w}^i$. The defect mark $n^i$ is computed based on the accuracy of $\mathbf{w}^i$ over a fixed validation dataset $\mathcal{D}_V$ on the server:
\begin{align}
    n^i = 1-\text{acc}(\mathbf{w}^i, \mathcal{D}_V),
\end{align}
where $\text{acc}(\mathbf{w}^i, \mathcal{D}_V)$ refers to the accuracy of testing $\mathbf{w}^i$ on the dataset $\mathcal{D}_V$. Then, we compute the quality evaluation marks $\hat{\mathbf{N}}=(\hat{n}^1,...,\hat{n}^N)$ as:
\begin{align}
    \hat{n}^i = f_{\text{QE}}(\mathbf{e}^i),
\end{align}
where $\hat{n}^i$ is the quality prediction of the $\mathbf{w}^i$, and $f_{\text{QE}}$ is the quality evaluation module composed of two FC layers.

Next, we compare $\mathbf{N}$ with $\hat{\mathbf{N}}$. After getting $\hat{\mathbf{W}}=(\hat{\mathbf{w}}^1,...,\hat{\mathbf{w}}^N)$, we use MSE loss function to compute $\hat{l}_1$ and $\hat{l}_2$:
\begin{align}
    \hat{l}_1(\theta_{\text{QEEN}}) &= \text{MSELoss}(\hat{\mathbf{W}}, \mathbf{W})\notag \\ 
    &= \frac{1}{|\mathbf{w}|N} \sum^N_{i=1}\sum^{|\mathbf{w}|}_{j=1}(\hat{w}^{\{i,j\}}-w^{\{i,j\}})^2,\\
    \hat{l}_2(\theta_{\text{QEEN}}) &= \text{MSELoss}(\hat{\mathbf{N}}, \mathbf{N}) = \frac{1}{N} \sum^{N}_{i=1}(\hat{n}^i-n^i)^2,
\end{align}
where $\theta_{\text{QEEN}}$ is the QEEN parameter, $|\mathbf{w}|$ is the number of uploaded model parameters, and $w^{\{i,j\}}$ is the $j$th parameter of the $i$th model.

Finally, we set different weights $\lambda^{\text{QEEN}}_1$ and $\lambda^{\text{QEEN}}_2$ for $\hat{l}_1$ and $\hat{l}_2$, respectively, to update the QEEN parameter $\theta_{\text{QEEN}}$ using joint gradient descent.

\subsection{Optimal Weight Assignment via DRL}\label{SAC in FL}


\subsubsection{MDP Modeling}
The process of the FL-based STLF can be modeled as an MDP. Assume the PP has a target mean absolute percentage error (MAPE) $\bar{\delta}$ which is usually close to $0$. At each round, $K$ UCs are randomly selected to conduct local training on their own datasets and upload model parameters to the PP. After receiving uploaded information as $\mathbf{s}$, the DRL model outputs $\mathbf{a}$, which is composed of weights to be assigned to all uploaded models. The details and explanations of $\mathcal{S}$, $\mathcal{A}$, and $\mathcal{R}$ of the MDP are defined as follows:

\textbf{State} $\mathcal{S}$: At round $t$, the state $\mathbf{s}_t$ is denoted as a vector $(\mathbf{e}^g_t, \mathbf{e}^1_t,..., \mathbf{e}^K_t, l^1_t, ..., l^K_t, \mathbf{a}_{t-1})$, where $\mathbf{e}^i_t$ denotes the embedding vector of model parameters of $C_i$, $\mathbf{e}^g_t$ denotes the embedding vector of the server's model parameters, $l^i_t$ denotes the local training loss of $C_i$, and $\mathbf{a}_{t-1}$ denotes the action at the previous round.

\textbf{Action} $\mathcal{A}$: The action, denoted as $\mathbf{a}_t=(a^1_t, a^2_t,..., a^K_t)$, is a weight vector, calculated by the DRL model, for a randomly selected subset of $K$ models at round $t$. All the weights in $\mathbf{a}_t$ are within $[0,1]$ and satisfy the constraint $\sum^{K}_{i=1}a^i_t = 1$. After obtaining the weight vectors, the server aggregates local model parameters to the global model parameters as follows:
\begin{equation}
\begin{aligned}
\mathbf{w}^g_t = \mathbf{a}_t \mathbf{W}_t = \sum^{K}_{i=1} a^i_t \mathbf{w}^i_t.
\end{aligned}
\end{equation}


\textbf{Reward} $\mathcal{R}$: At round $t$, the current reward $r_t$ guides $\mathbf{a}_{t+1}$ to maximize the cumulative return $G$, i.e. the goal of DRL. We design a compound reward by combining two sub-rewards with appropriate weights $\beta_i$ and a discount factor $\gamma$, which can be formulated as:
\begin{align}
G &= \sum^{T}_{t=1}\gamma^{t-1} (\beta_1 r^1_t+\beta_2 r^2_t).\\
\label{r1} r^1_t&= \kappa^{\bar{\delta}-\delta_t} - 1, \\
\label{r2} r^2_t&= - \frac{1}{K} \sum^{K}_{i=1} {(\bar{n}^i_t - a^i_t)}^2.
\end{align}

In Eq. (\ref{r1}), $r^1_t$ aims to minimize the global model's MAPE. The exponential term $\bar{\delta}-\delta_t$ represents the MAPE gap, where $\delta_t$, which is usually within $[0,1]$, is the global model's MAPE on the held-out validation dataset at round $t$. To mitigate the slow convergence caused by diminishing marginal effect \cite{wang2020optimizing}, we use $\kappa$, a positive constant, to ensure an exponential growth of $r^1_t$. Assume $\bar{\delta}$ is $0$, thus $\bar{\delta}-\delta_t$ is within $[-1,0]$, and the term $\kappa^{\bar{\delta}-\delta_t}$ is within $[\frac{1}{\kappa},1]$. As the more rounds the agent takes, the less cumulative reward the agent obtains, we need to punish the agent for finishing training in more rounds. Therefore, the second term $-1$ is used as the time penalty at each round $t$ to set $r^1_t$ within $[\frac{1}{\kappa}-1,0]$ for faster convergence.

Eq. (\ref{r2}) aims to provide auxiliary information for the agent to reduce exploration time. After obtaining the quality prediction mark $\hat{n}^i_t$ from QEEN, we normalize $\hat{n}^i_t$ to $\bar{n}^i_t$ within $[0,1]$ to calculate the MSE loss of $\bar{n}^i_t$ and $a^i_t$, where $\sum^K_{i=1}\bar{n}^i_t=1$. Similar to Eq. (\ref{r1}), Eq. (\ref{r2}) is set to negative for the time penalty.


\subsubsection{Soft Actor-Critic}

We adopt SAC \cite{haarnoja2018soft} to solve the MDP. At the end of each round $t$, the tuple $(\mathbf{s}_t,\mathbf{a}_t,r_t,\mathbf{s}_{t+1})$, which is denoted as $\tau$, is recorded in the replay buffer $\mathcal{B}$.



In SAC, the action $\mathbf{a}$ is sampled from a Gaussian distribution $\mathcal{N}(\mu,\sigma^2)$, whose mean $\mu$ and standard deviation $\sigma$ are outputted from the actor network $\pi$ of the DRL model:
\begin{align}
    \mu, \sigma = \pi(\cdot|\mathbf{s}).
\end{align}

For each iteration, SAC samples a batch of $\tau$ from $\mathcal{B}$ and updates the DRL network parameters. To deal with poor sampling efficiency and data unbalance in DRL \cite{sutton2018reinforcement}, we adopt two techniques of replay buffer named emphasizing recent experience (ERE) \cite{wang2019boosting} and prioritized experience replay (PER) \cite{schaul2015prioritized} to sample data with priority and emphasis. 


The update procedure of SAC is shown in Fig. \ref{DearFSAC}, where $\mathbf{a}'$ is the action of next time step, i.e. the next FL round, and $\dot{\mathbf{a}}$ is the action of last time step, i.e. the last FL round.

\subsection{Workflow of DearFSAC}
The workflow of the DearFSAC approach is summarized in Algorithm \ref{alg3}, and are described as follows.
\begin{itemize}
    \item At the first communication round, the global model parameters are initialized on the PP, denoted as $\mathbf{w}^g_{\text{Init}}$ (line $3$).
    \item All UCs send download request to the PP and download the latest global STLF model as the local STLF model. After local training, all UCs obtained corresponding updated models and training loss. During the local training, defects may occur in datasets and model parameters (lines $5-11$).
    \item The PP selects $K$ UCs to upload models and training loss. After obtaining embedding vectors by inputting model parameters into QEEN, the PP concatenate embedding vectors, training loss, and the action at last round to get the state. By inputting the state into the DRL model, the mean $\mu$ and variance $\sigma$ of the Gaussian distribution \cite{haarnoja2018soft} are obtained to sample current action $\mathbf{a}$. Finally, by multiplying $\mathbf{a}$ and $\mathbf{W}$, the PP obtains the global model $\mathbf{w}^g$ and send it to selected UCs. The whole process loops until convergence (line $15-24$).
\end{itemize}


\begin{algorithm}[H]
\footnotesize
\caption{Workflow of the DearFSAC Approach}
\begin{algorithmic}[1]
\STATE \textbf{Input:} Initial global model $\mathbf{w}^g_{\text{Init}}$; the set of UCs $C=\{C_1,...,C_N\}$; the number of selection $K$; the learning rate of local training $\alpha_{\text{LSTM}}$; initial action $\mathbf{a}_{\text{Init}}$; data owned by each UCs $\{\mathcal{D}_1,...,\mathcal{D}_N\}$, where $\mathcal{D}_i=<\mathbf{X}_i,\mathbf{y}_i>$; empty prioritized experience replay buffer $\mathcal{B}$.
\STATE \textbf{Output:} The global STLF model $\mathbf{w}^g$.

\STATE Initial $\mathbf{w}^g_{\text{Init}}$ and $\mathbf{a}_{\text{Init}}$ for the PP;
\STATE \textbf{for} each FL iteration $t$ \textbf{do}
\STATE \hspace{0.5cm} $N$ \textbf{UCs execute:}
\STATE \hspace{0.5cm} \textbf{for} each local model $\mathbf{w}^i_{t}$ owned by $C_i$ \textbf{do}
\STATE \hspace{0.5cm} \hspace{0.5cm} $\mathbf{w}^i_t \gets \mathbf{w}^g_t$, if $t=0$ then $\mathbf{w}^g_{\text{Init}}$;

\STATE \hspace{0.5cm} \hspace{0.5cm} $\hat{\mathbf{y}}_i \gets \text{STLF}(\mathbf{w}^i_t, \mathbf{X}_i)$;
\STATE \hspace{0.5cm} \hspace{0.5cm}
$l^i_t \gets \text{MSELoss}(\hat{\mathbf{y}}_i,\mathbf{y}_i)$;
\STATE \hspace{0.5cm} \hspace{0.5cm}
$\mathbf{w}^i_t \gets \mathbf{w}^i_t - \alpha_{\text{LSTM}} \bigtriangledown l^i_t$;

\STATE \hspace{0.5cm} \hspace{0.5cm} If selected, upload $\mathbf{w}^i_t$ and $l^i_t$ to the PP;
\STATE \hspace{0.5cm} \textbf{end for}
\STATE
\STATE \hspace{0.5cm} \textbf{PP executes:}
\STATE \hspace{0.5cm} Select $K$ UCs $\{ C_{k_1},...,C_{k_K}\}$ to upload;
\STATE \hspace{0.5cm} Reset $\mathbf{W}_t$, $\mathbf{L}_t$, and $\mathbf{E}_t$ to record $\mathbf{w}_t$, $l_t$, and $\mathbf{e}_t$ respectively;
\STATE \hspace{0.5cm} $\mathbf{e}^g_t \gets f_{\text{Enc}}(\theta_{\text{QEEN}},\mathbf{w}^g_t)$;
\STATE \hspace{0.5cm} \textbf{for} each selected UC $C^{k_j}$ \textbf{do}
\STATE \hspace{0.5cm}\hspace{0.5cm} $\mathbf{e}^{k_j}_t \gets f_{\text{Enc}}(\theta_{\text{QEEN}},\mathbf{w}^{k_j}_t)$;
\STATE \hspace{0.5cm}\hspace{0.5cm} $\mathbf{W}_t \gets \mathbf{W}_t \cup \{\mathbf{w}^{k_j}_t\}, \mathbf{L}_t \gets \mathbf{L}_t \cup \{l^{k_j}_t\}, \mathbf{E}_t \gets \mathbf{E}_t \cup \{\mathbf{e}^{k_j}_t\}$;
\STATE \hspace{0.5cm} \textbf{end for}
\STATE \hspace{0.5cm} $\mathbf{s}_t \gets (\mathbf{e}^g_t, \mathbf{E}_t, \mathbf{L}_t, \mathbf{a}_{t-1})$, if $t=0$ then $(\mathbf{e}^g_t, \mathbf{E}_t, \mathbf{L}_t, \mathbf{a}_{\text{Init}})$;
\STATE \hspace{0.5cm} $\mathbf{w}^g_{t+1} \gets \mathbf{a}_t \mathbf{W}_t, \mathbf{a}_t \sim \mathcal{N}(\mu,\sigma^2)$, where $\mu, \sigma \gets \pi(\cdot|\mathbf{s})$;
\STATE \textbf{end for}
\end{algorithmic}
\label{alg3}
\end{algorithm}

\section{Simulations \& Analysis}\label{sec5}

\subsection{Simulation Setup}\label{setup}
\subsubsection{Datasets}
We conduct simulations using data from Nuuka open API \cite{nuuka}, containing the basic information and energy data of Helsinki’s utility and service properties. As mention in Section \ref{sec3}, the aim of this work is to obtain an accurate STLF model using FL framework with defects occurring. To alleviate affect of different characteristics, we use K-means to cluster. After data cleansing and clustering, data of $100$ UCs spanning $2$ years from $1$st March $2017$ to $1$st March $2019$ are used in this work. Each UC is set as a client. All of the clients have hourly resolution data. Considering the seasonal factors, we further split the dataset into four seasons. Then, the data before the last week of each season is used for training, and the last one week of each season is set as the test dataset.



\subsubsection{Defect Types}
As mentioned above, the proportion of defective models is $p_M$. Then we design $4$ scenarios in Table II based on $3$ types of defects as follows:
\begin{itemize}
    \item \textbf{DIAs:} Before making sliding windows, we simulate a normally-distributed DIAs on the training dataset by randomly selecting $k\%$ of all data points to alter their loads by multiplying $1+p\%$, where $p$ is sampled from a Gaussian distribution $\mathcal{N}_a$ with mean $\mu_a$ and standard deviation $\sigma_a$ \cite{luo2018benchmarking}.
    \item \textbf{Communication noises:} We apply the signal to noise ratio (SNR) to model the noises, and the modified parameters $\tilde{\mathbf{w}}$ is calculated as follows:
    \begin{align}
        \tilde{\mathbf{w}}=\mathbf{w}+\frac{\mathbf{w}}{10^{\text{SNR}_{\text{dB}}/10}},
    \end{align}
    where $\text{SNR}_{\text{dB}}$ indicates the level of noises.
    \item \textbf{Mixed defect:} Consider more severe conditions, we design the mixed defect by adding both DIAs and communication noises into the same client.
\end{itemize}


\subsubsection{Comparison Approaches}
To evaluate the performance of our approach, $3$ approaches are adopted for comparison:
\begin{itemize}
    \item \textit{Centralized learning (CL-LSTM)}: The data is gathered to train a global model. Note that the data privacy may be intruded in this learning framework. Based on CL, we can construct the LSTM as CL-LSTM. 
    \item \textit{Federated learning (FL-LSTM)}: As the most common approach in FL \cite{mcmahan2017communication}, \textbf{federated averaged (FedAvg)}, which assigns averaged weights to local models, is adopted based on the proposed LSTM. 
    \item \textit{DearFSAC without QEEN (FL-LSTM-SAC)}: QEEN is an auto-encoder that needs extra computational costs for training. Besides, during FL communication, the process of QEEN may spend more time. Therefore, FL-LSTM-SAC is adopted to evaluate the necessity of QEEN in our proposed approach.
\end{itemize}


\subsubsection{Evaluation Metrics}
The following $2$ metrics are used to evaluate the performance of our approach on the test dataset:
\begin{itemize}
    \item MAPE is a percentage quantifying the size of the prediction error, which is defined as:
    \begin{align}
        \text{MAPE}= \frac{100\%}{\hat{N}} \sum^{\hat{N}}_{i=1}{|\frac{y^i-\hat{y}^i}{y^i}|},
    \end{align}
    where $y^i$ is the actual value, $\hat{y}^i$ is the predicted value, and $\hat{N}$ is the number of predicted values.
    \item The root mean square error (RMSE) quantifies the error in terms of electricity, which is defined as:
    \begin{align}
        \text{RMSE}= \sqrt{\frac{\sum^{\hat{N}}_{i=1}(y^i-\hat{y}^i)^2}{\hat{N}}}.
    \end{align}
\end{itemize}

To evaluate the performance of approaches more precisely, we conduct simulations for $5$ times and compute the averaging metrics.
\begin{table}[]
\centering
\caption{$4$ scenarios of FL-based STLF training in wholesale markets}
\setlength{\tabcolsep}{12mm}{
\begin{tabular}{cc}
\hline\hline
Scenario & Defect type   \\ \hline
I        & None         \\
II        & Data integrity attacks       \\
III       & Communication noises        \\
IV      & Mixed defect    \\ \hline\hline
\end{tabular}}
\end{table}

\begin{table}[]
\centering
\caption{Hyperparameters of the proposed approach}
\renewcommand\arraystretch{1.4} 
\setlength{\tabcolsep}{5mm}{
\begin{tabular}{ccc}
\hline\hline
\multicolumn{1}{l}{}                      & Parameter description & Value     \\ \hline
\multicolumn{1}{l}{\multirow{2}{*}{FL}} & Total training round           & 1000       \\
\multicolumn{1}{l}{}                      & Local training epoch         & 1      \\ \hline
\multicolumn{1}{l}{\multirow{5}{*}{LSTM}} & Hidden layer size           & 512       \\
\multicolumn{1}{l}{}                      & Learning rate         & 1e-4      \\
\multicolumn{1}{l}{}                      & Length of sliding windows $l_{\text{back}}$                 & 24        \\
\multicolumn{1}{l}{}                      & Interval of sliding windows $l_{\text{ahead}}$                & 1         \\
\multicolumn{1}{l}{}                      & Optimizer             & Adam      \\ \hline
\multirow{7}{*}{SAC}                      & Target MAPE $\bar{\delta}$                 & 2\%       \\
                                          & $r^1$'s base number $\kappa$                 & 64        \\
                                          & Reward weight set $(\beta_1,\beta_2)$         & (0.5,0.5) \\
                                          & Decay rate $\gamma$                 & 0.99      \\
                                          & Buffer size $|\mathcal{B}|$          & 1e5       \\
                                          & Soft update rate $\rho$                   & 5e-3      \\
                                          & Learning rate         & 3e-4      \\ \hline
QEEN                                      & Loss weight set $(\lambda^{\text{QEEN}}_1,\lambda^{\text{QEEN}}_2)$           & (0.5,0.5) \\\hline\hline
\end{tabular}}
\end{table}

\subsubsection{Details of Setup}
All simulations and approaches are coded in Python. DearFSAC and the above comparison approaches are implemented using Pytorch, and conducted on a personal computer with an NVIDIA GeForce RTX 2080 Ti GPU. Referring to \cite{kong2017short, haarnoja2018soft}, the hyperparameters of the proposed DearFSAC are listed in Table III.

\subsection{Performance Analysis}


In this subsection, $100$ UCs participate in FL training under $4$ scenarios, in which $p_K$ is set as $10\%$, and $p_M$ is set as $20\%$ for Scenario II, III, and IV. For DIAs, we set $k$, $\mu_a$, and $\sigma_a$ as $30\%$, $30$, and $50$, respectively. For communication noises, we set $\text{SNR}_{\text{dB}}$ as $30$. 

Through comparisons, the $4$ approaches are evaluated in terms of MAPE and RMSE in Table IV, where the best results under each scenario are in bold. Note that there is a $20\%$ probability of adding defects into the CL model in each epoch.

As shown in Table IV, the results under Scenario I show that the MAPE and RMSE of DearFSAC are slightly lower than the other $2$ FL-based approaches and much lower than CL-LSTM. However, under Scenario II, III, and IV, DearFSAC maintains almost the same performance as Scenario I and outperforms the other $3$ approaches. The reasons lie in $2$ key points: (1) When no defects occur, nearly averaged weights can conduct feasible model aggregation and obtain an accurate global model because of the K-means clustering. CL-LSTM is a general model trained on the whole dataset and lacks personalization. However, FL-based approaches use model aggregation to share parameters for preventing overfitting. Since the FL training process is not a convex problem, the distributed training manner may achieve a better performance when better sub-optimal solutions are found. (2) When defects reduce the model quality, errors are accumulated in the global model during the FL process. Due to DearFSAC's capability of assigning nearly optimal weights to uploaded models, only DearFSAC can effectively conduct model aggregation for each FL communication round. 

\begin{figure} \centering    
\subfigure {
\includegraphics[width=0.46\textwidth]{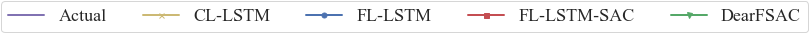}  }
\subfigure {
\includegraphics[width=0.22\textwidth]{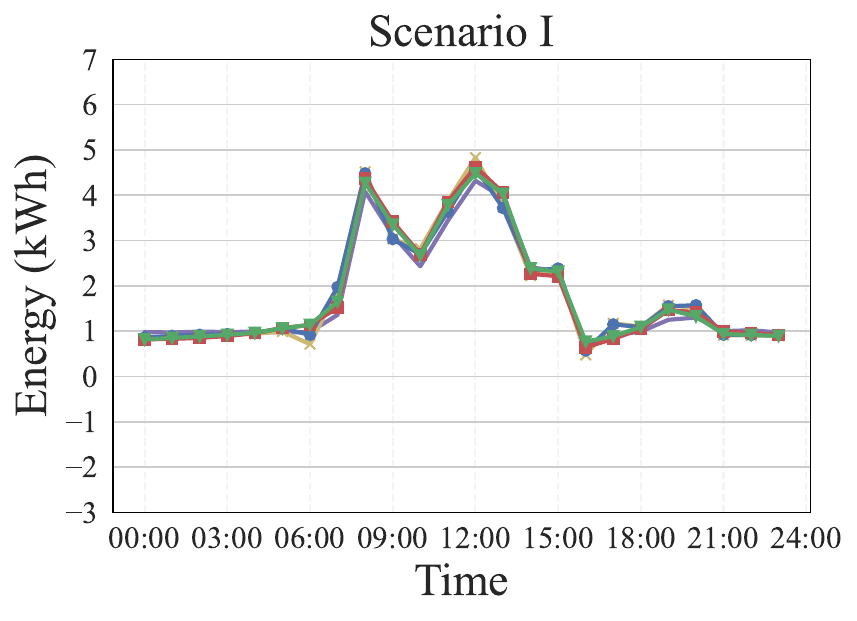}  }
\subfigure { 
\includegraphics[width=0.22\textwidth]{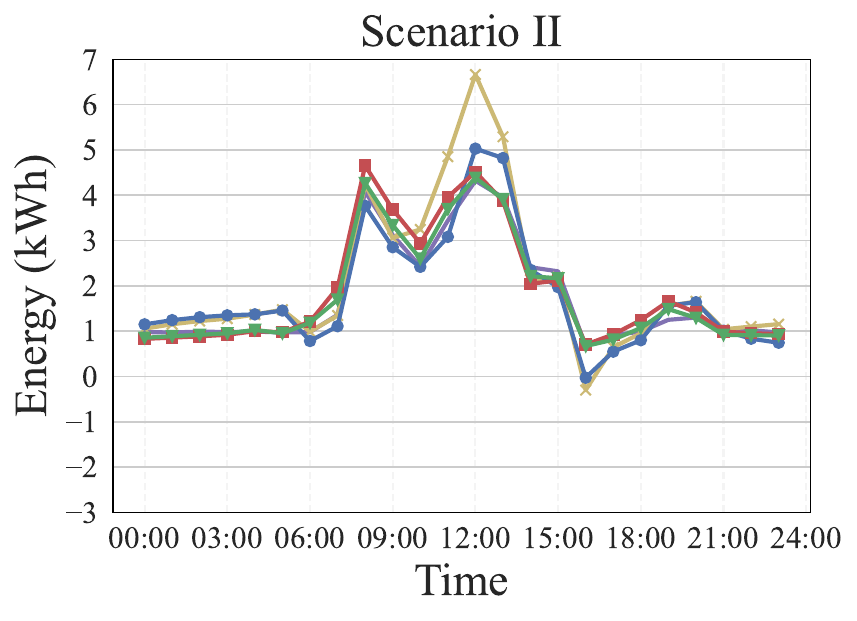}  }\\
\subfigure { \includegraphics[width=0.22\textwidth]{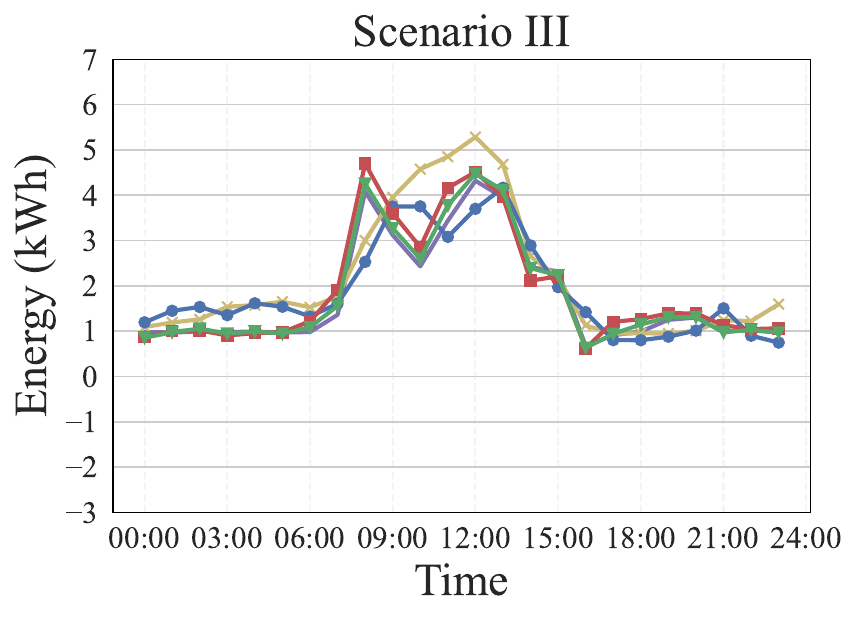}    
}  
\subfigure { \includegraphics[width=0.22\textwidth]{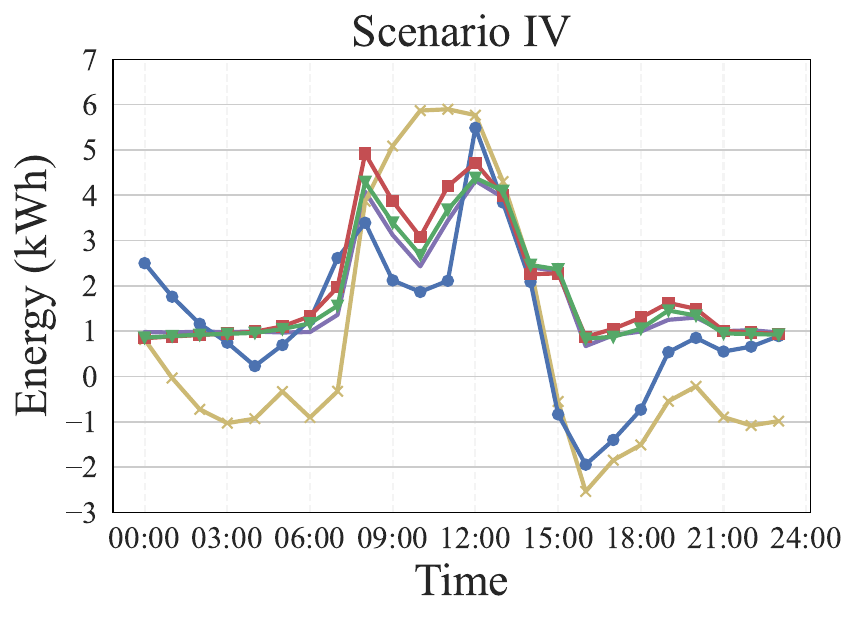}    
}  
\caption{$4$ approaches' hourly forecasting on 1721 Lpk Aleksi ja Dh Alexia utility's load on $5$th February under $4$ scenarios.}
\label{forecast curve}
\end{figure}

The load forecasting curves of $4$ approaches under $4$ scenarios on $5$th February are shown in Fig. \ref{forecast curve}. We can see that though the actual demand fluctuates, $4$ approaches under Scenario I still forecast accurately, indicating that our LSTM-based STLF model is effective on this dataset. Under Scenario II, III, and IV, FL-LSTM-SAC performs similarly well, showing the effectiveness of the DRL algorithm. Even so, DearFSAC performs better than FL-LSTM-SAC, proving the superiority of our proposed approach. On the contrary, the performance of the other two approaches is not so good. When DIAs occur, as shown in Scenario II, the forecasting results of CL-LSTM and FL-LSTM deviate from the actual demand, but the general trend is similar to the one of actual demand, indicating that DIAs can only affect the individual values of forecasting and have little impact on the tendency. Nevertheless, as shown in Scenario III, communication noises are seemingly capable of affecting the forecasting tendency of CL-LSTM, while the one of FL-LSTM is more stable. More severely, the mixed defects significantly harm the STLF performance of CL-LSTM and FL-LSTM, where FL-LSTM still seems to have a general trend similar to the actual demand. The reason is that through model aggregation, the global model can alleviate the impact of communication noises, whereas the errors are also accumulated in each FL communication round if defective models obtain respectable weights during model aggregation.

\begin{table*}[]
\centering
\caption{Comparison of performance using different approaches under $4$ scenarios}
\setlength{\tabcolsep}{2mm}{
\begin{tabular}{ccccccccc}
\hline\hline
                          & \multicolumn{4}{c}{RMSE (kW)}                                                                                                 & \multicolumn{4}{c}{MAPE (\%)}                                                                                                 \\
                          & \multicolumn{1}{c}{Scenario I} & \multicolumn{1}{c}{Scenario II} & \multicolumn{1}{c}{Scenario III} & \multicolumn{1}{c}{Scenario IV} & \multicolumn{1}{c}{Scenario I} & \multicolumn{1}{c}{Scenario II} & \multicolumn{1}{c}{Scenario III} & \multicolumn{1}{c}{Scenario IV} \\ \hline
CL-LSTM                   & $2.62\pm0.25$                              & $3.75 \pm 0.23$                              & $3.87 \pm 1.55$                              & $13.54\pm4.61$                              & $1.24 \pm 0.07$                              & $1.85\pm0.09$                              & $1.87\pm0.74$                              & $7.22\pm1.53$                              \\
FL-LSTM           & $2.12 \pm 0.16$                              & $3.26\pm0.21$                              & $3.25 \pm 0.78$                              & $12.77\pm3.14$                              & $1.05\pm0.08$                              & $1.79\pm0.13$                              & $1.77\pm0.53$                              & $6.69\pm1.07$                              \\
FL-LSTM-SAC               & $2.14 \pm 0.35$                              & $2.23 \pm 0.41$                              & $2.74 \pm 0.85$                              & $2.95 \pm 0.97$                              & $1.04 \pm 0.15$                              & $1.05 \pm 0.11$                              & $1.45 \pm 0.39$                              & $1.66 \pm 0.51$                              \\
\textbf{DearFSAC} & $\mathbf{1.85}\pm\mathbf{0.09}$                              & $\mathbf{1.87}\pm\mathbf{0.09}$                              & $\mathbf{1.86}\pm\mathbf{0.12}$                              & $\mathbf{1.88}\pm\mathbf{0.13}$                              & $\mathbf{0.98}\pm\mathbf{0.03}$                              & $\mathbf{0.99}\pm\mathbf{0.03}$                              & $\mathbf{0.99}\pm\mathbf{0.05}$                              & $\mathbf{1.01}\pm\mathbf{0.06}$                              \\ \hline\hline
\end{tabular}}
\end{table*}

\begin{table}[]
\centering
\caption{Runtime for $1000$-round training under Scenario IV}
\setlength{\tabcolsep}{12mm}{
\begin{tabular}{cc}
\hline\hline
            & Runtime (second) \\ \hline
CL-LSTM          &  $2624.9$                \\
FL-LSTM  &      $1057.4$            \\
FL-LSTM-SAC &       $1665.8$           \\
\textbf{DearFSAC} &     $\mathbf{1544.3}$             \\ \hline\hline
\end{tabular}
}
\end{table}

Then, to assess the computational efficiency, runtime for the $1000$-round FL training is compared, where CL-LSTM conducts training for $1000$ epochs. As shown in Table V, the runtime of FL-based approaches is much lower than that of CL-LSTM, because the FL-based framework executes parallel local training in each UC to update models in a distributed manner. Besides, though DRL-adopted approaches cost more time than FL-LSTM, the runtime of FL-LSTM-SAC and DearFSAC is still acceptable. Furthermore, compared with FL-LSTM-SAC, DearFSAC's runtime is even shorter. The reason is that directly inputting high-dimensional model parameters into the DRL model costs plenty of time, while QEEN spends a little time significantly reducing the model dimension for faster DRL computation. 

\begin{figure*} \centering   
\subfigure {
\includegraphics[width=0.3\textwidth]{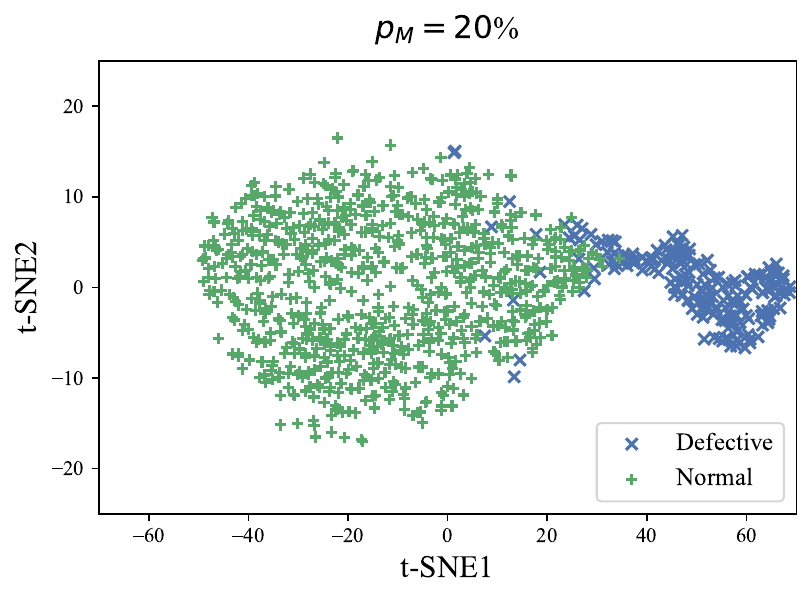}  }
\subfigure { 
\includegraphics[width=0.3\textwidth]{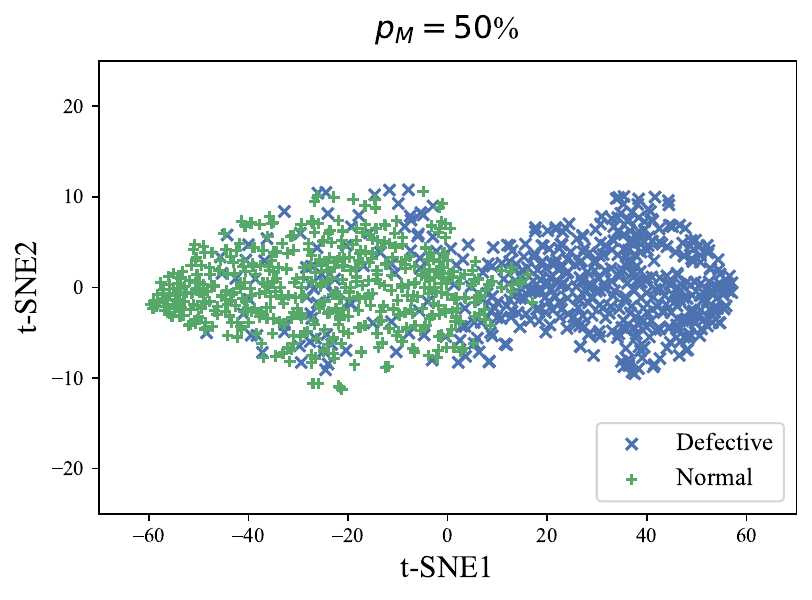}  }
\subfigure { \includegraphics[width=0.3\textwidth]{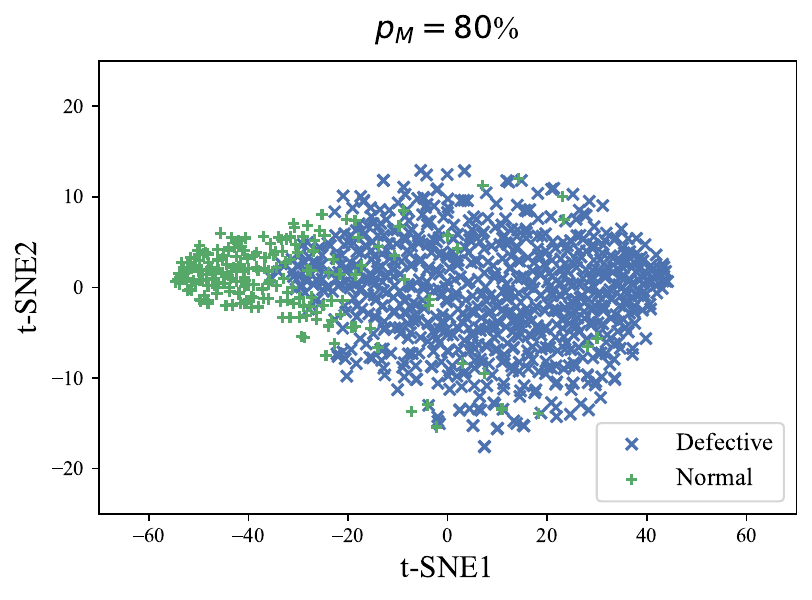}    }  
\caption{QEEN divides $1000$ model embedding vectors to $2$ clusters, where $p_M$ is set to $20$\%, $50$\%, and $80$\%.}
\label{tsne}
\end{figure*}

To prove that QEEN indeed outputs embedding vectors containing defect information, we adopt t-SNE to visualize each embedding vector in a two-dimension space, setting $N$ as $100$ and changing $p_M$. As shown in Fig. \ref{tsne}, $1000$ embedding vectors are divided into $2$ clusters by QEEN, where normal and defective models are labeled in advance, indicating that QEEN can disguise defective model parameters and normal model parameters effectively.

\subsection{Robustness Analysis}
In this subsection, we evaluate the robustness of DearFSAC by changing $p_M$ of Scenario II and III. Then, we change $k$, $\mu_a$, and $\sigma_a$ under Scenario II, and $\text{SNR}_{\text{dB}}$ under Scenario III. Note that we set $N$ and $p_K$ as $100$ and $10\%$ respectively. The proposed DearFSAC is compared with FL-LSTM.

\subsubsection{Different proportions of defective models}
\begin{figure} \centering    
\subfigure {
\includegraphics[width=0.22\textwidth]{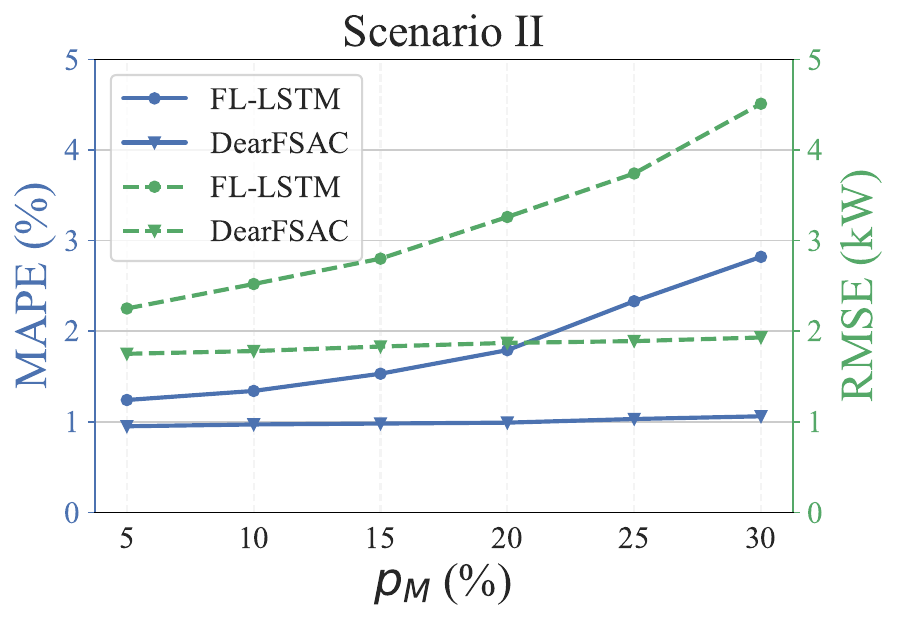}  }
\subfigure { 
\includegraphics[width=0.22\textwidth]{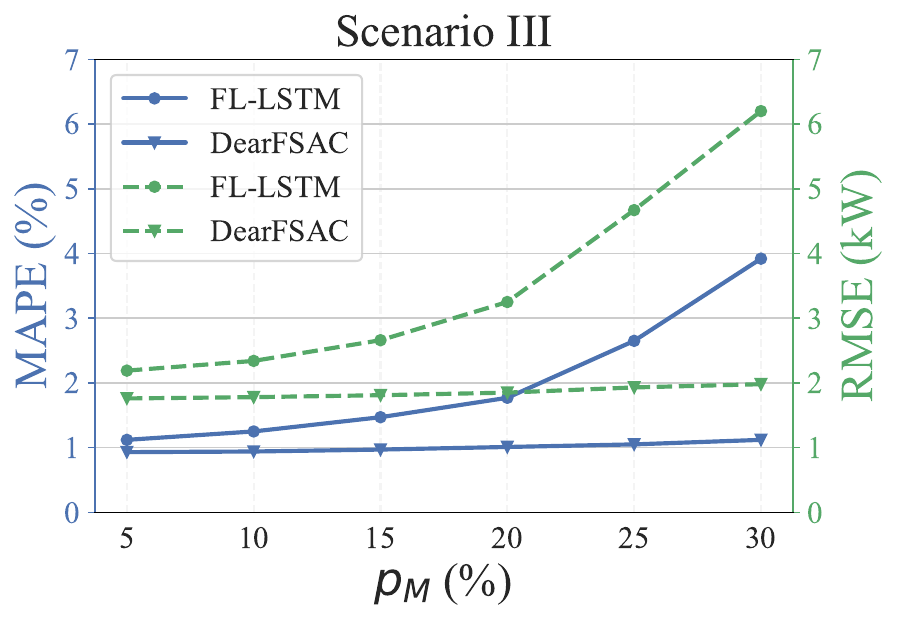}  }
\caption{MAPE and RMSE of FL-LSTM and DearFSAC with different $p_M$ under Scenario II and III.}
\label{defect M}
\end{figure}

Firstly, we conduct simulations under Scenario II and III while changing the value of $p_M$ to study how the proportion of defective models affects the performance. As shown in Fig. \ref{defect M}, as $p_M$ increases, the MAPE and RMSE of FL-LSTM increase dramatically, indicating that the larger the $p_M$ is, the worse the performance of FL-LSTM is. However, DearFSAC performs well despite the value of $p_M$. The reason is that if $p_M$ is small, just a few defects affect the model aggregation, which can be covered by other normal model parameters in each round. Nevertheless, if $p_M$ grows, errors in each round will be increasingly accumulated. Rather than cover defects, model aggregation will be deteriorated by defective models and output a worse and worse global model which is seriously harmful to UCs' local models. In addition, we can conclude that $p_M$ influences DIAs and communication noises similarly in this paper.

\subsubsection{Different levels of DIAs}
To simulate different levels of DIAs, we adjust $k$, $\mu_a$, and $\sigma_a$ under Scenario II. Then we compare MAPE of FL-LSTM and DearFSAC. 

\begin{figure} \centering    
\subfigure {
\includegraphics[width=0.22\textwidth]{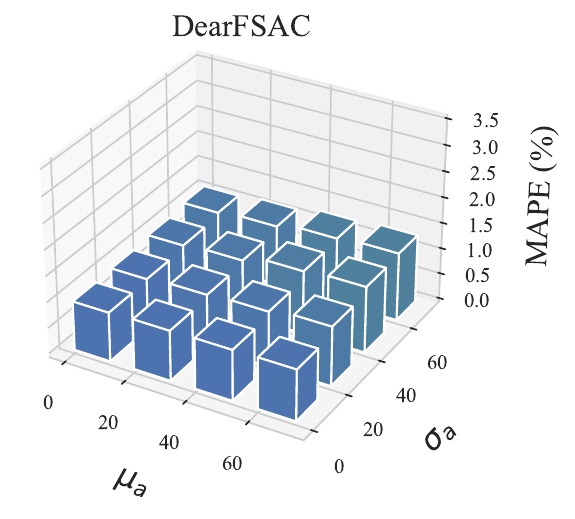}  }
\subfigure { 
\includegraphics[width=0.22\textwidth]{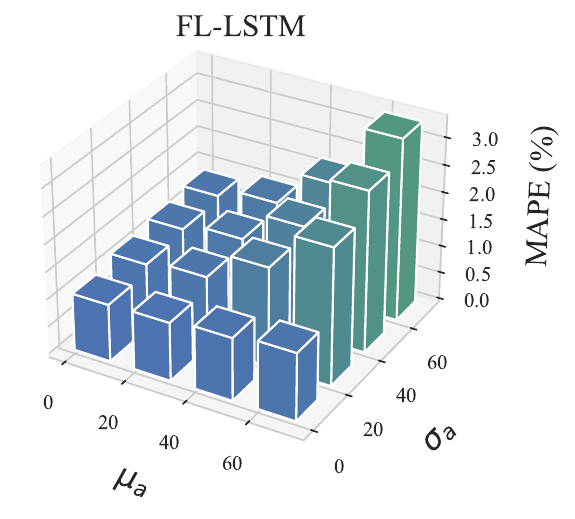}  }
\caption{MAPE of FL-LSTM and DearFSAC with different $\mu_a$ and $\sigma_a$ under Scenario II.}
\label{DIA_mu}
\end{figure}

As shown in Fig. \ref{DIA_mu}, we set both the values of $\mu_a$ and $\sigma_a$ to be between $0$ and $60$ by increments of $20$. We can see that when $\mu_a$ or $\sigma_a$ increases, the MAPE of FL-LSTM gets larger, indicating that the larger the DIA level is, the harder the defects are. On the contrary, our proposed approach maintains good performance, which just gets worse slightly. The reason is that DearFSAC can recognize DIAs based on QEEN and assign lower weights to models with stronger DIAs. Besides, the variation of $\mu_a$ affect the model performance more seriously with larger $\sigma_a$, showing that $\sigma_a$ is the more dominant factor in DIAs.

\begin{figure} \centering    
\subfigure {
\includegraphics[width=0.23\textwidth]{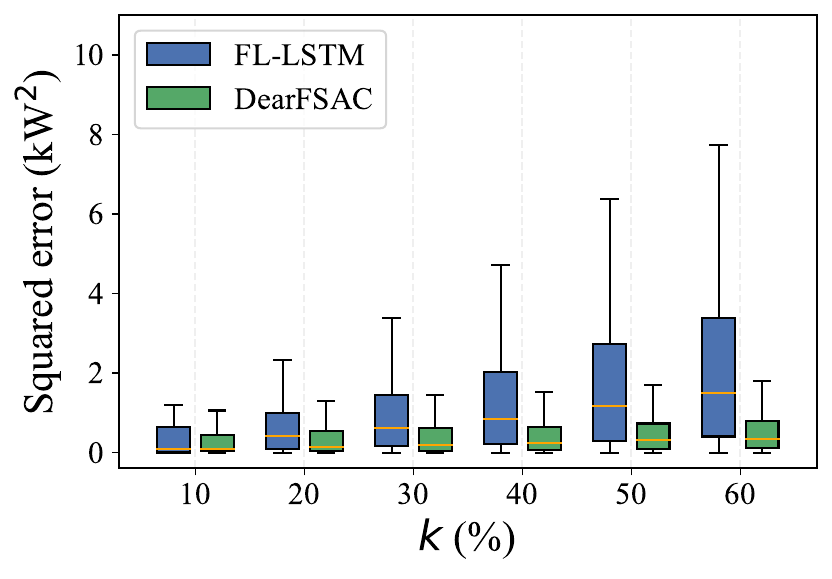}  }
\subfigure { 
\includegraphics[width=0.22\textwidth]{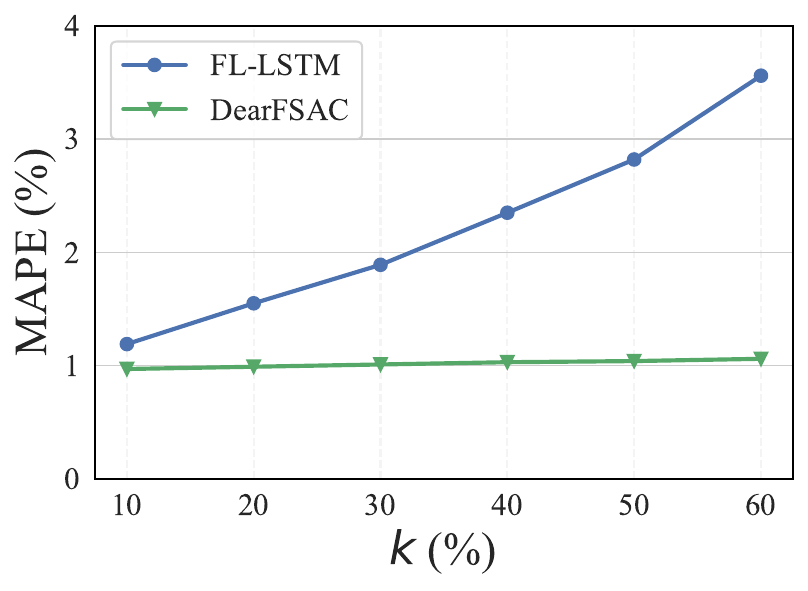}  }
\caption{Squared errors and MAPE comparison of FL-LSTM and DearFSAC with different $k$ under Scenario II.}
\label{k}
\end{figure}

Furthermore, we change $k$ to study how the proportion of data points under DIAs affects the global model performance. In addition to MAPE, the specific squared error of each point is also recorded for more clear observation. Fig. \ref{k} shows that as the $k$ increases, there is a steady growth of the MAPE of FL-LSTM, while DearFSAC's MAPE keeps stable. The observation enhances the conclusion that DearFSAC has the capability of recognizing models with different levels of DIAs.


\begin{figure} \centering    
\subfigure {
\includegraphics[width=0.23\textwidth]{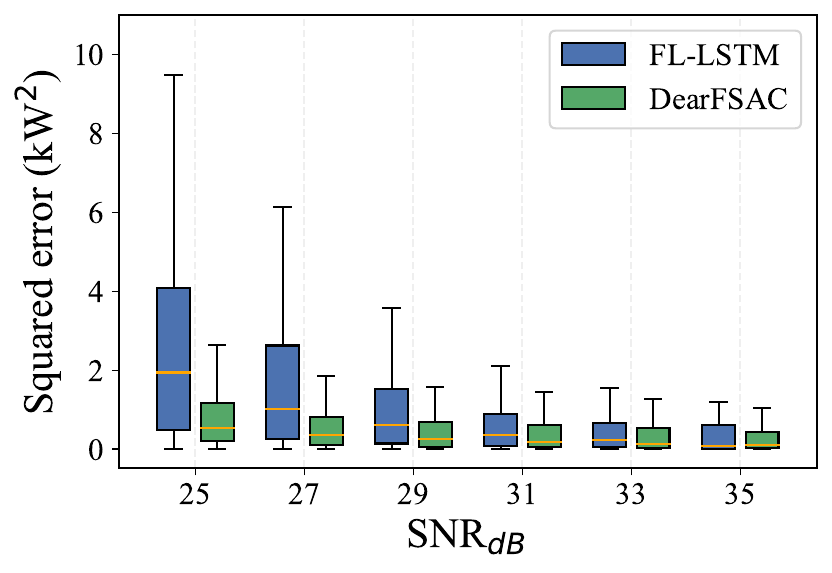}  }
\subfigure { 
\includegraphics[width=0.22\textwidth]{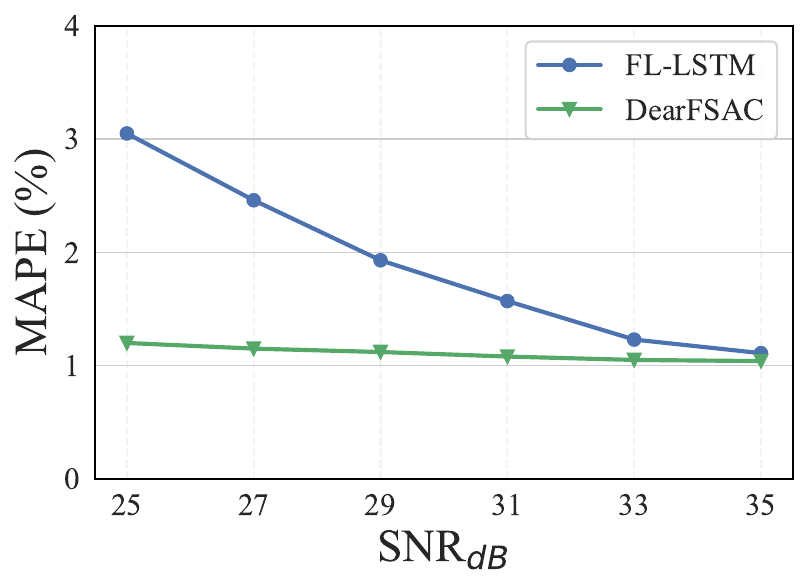}  }
\caption{Squared errors and MAPE comparison of FL-LSTM and DearFSAC with different $\text{SNR}_{\text{dB}}$ under Scenario III.}
\label{snrdb}
\end{figure}

\subsubsection{Different levels of communication noises}
To simulate different levels of communication noises, we adjust $\text{SNR}_{\text{dB}}$ under Scenario III. Then we compare the performance of FL-LSTM and DearFSAC. As shown in Fig. \ref{snrdb}, the performance of FL-LSTM increasingly reduces as the $\text{SNR}_{\text{dB}}$ increases, while the one of DearFSAC just increases a little and still keeps in a low range. The reason is that the smaller $\text{SNR}_{\text{dB}}$ is, the more severe the communication noises are. When models are extremely defective, DearFSAC can assign reasonably low weights to them while FL-LSTM assigns averaged weights and introduces large errors to the global model. Thus, DearFSAC is capable of holding a feasible performance when the communication noises are serious.

\subsection{Scalability Analysis}
\begin{figure} \centering    
\subfigure {
\includegraphics[width=0.22\textwidth]{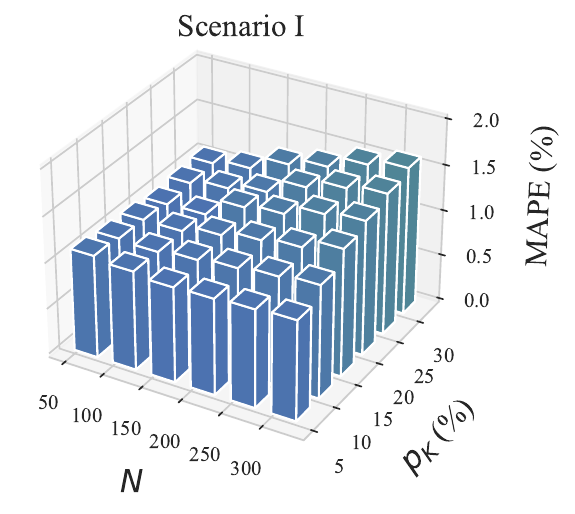}  }
\subfigure { 
\includegraphics[width=0.22\textwidth]{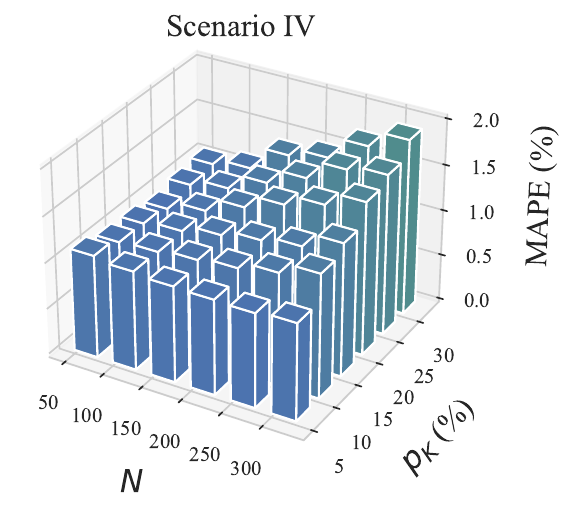}  }
\caption{MAPE of DearFSAC with different $p_K$ and $N$ under Scenario I and Scenario IV.}
\label{scala}
\end{figure}

In this subsection, we evaluate the scalability of DearFSAC by changing total client number $N$ and selecting proportion $p_K$ under Scenario I and IV. We set $k$, $\mu_a$, $\sigma_a$, and $\text{SNR}_{\text{dB}}$ as $30\%$, $30$, $50$, and $30$ respectively. As shown in Fig. \ref{scala}, small variation occurs in performance when $p_K$ and $N$ are both small. However, if fixing $N$, the performance drops with $p_K$ increasing, and the same phenomenon applies to $p_K$, showing that the number of UCs and selection proportion of uploaded models should be both kept within a small range. Besides, compared with the performance under Scenario I, our proposed approach has a slightly worse performance under Scenario IV. The reason is that increasing number of models to be aggregated leads to larger computational costs. As DearFSAC can not assign perfect weights to all selected models, accumulated errors cause a larger probability of introducing defects into the global model. Note that when $N$ is $50$, small $p_K$ makes the performance a little worse, indicating that too few models participating in the aggregation will weaken the global model. In general, our proposed approach is scalable when changing the total number of UCs and the selection proportion of uploaded models.

\section{Conclusion \& Future Works}\label{sec6}
In this paper, DearFSAC, a DRL-assisted FL approach, is proposed to robustly integrate STLF models for individual PPs. By adopting FL, one PP can obtain an accurate STLF model just using UCs' local models, which protects data privacy. Considering defects, the SAC algorithm is adopted to conduct robust model aggregation. In addition, for better convergence of our approach, QEEN, an auto-encoder, is designed for both dimension reduction and quality evaluation of uploaded models. In simulations, our approach shows the superiority and robustness of the proposed approach in utility demand forecasting.

In the future, we will focus on the following three aspects: 1) Stronger robustness considering more types of defects; 2) improvement under more distributed scenarios; 3) extension of more types of energy resources.

\maketitle

\small
\bibliographystyle{IEEEtran}
\bibliography{reference}
\end{document}